%% file: TOP-13-001_temp.tex
\pdfoutput=1

\documentclass[11pt,twoside,a4paper,cmspaper,final,collab]{cms-tdr}
\def\svnVersion{339860}\def\cmsCernNoTag{CERN-PH-EP/2015-282}\def\cmsCernDate{\today}\def\cmsMessage{Published in the Journal of High Energy Physics as \href{http://dx.doi.org/10.1007/JHEP04(2016)073}{\doi{10.1007/JHEP04(2016)073.}}}

\begin{document}\cmsNoteHeader{TOP-13-001}

\hyphenation{had-ron-i-za-tion}
\hyphenation{cal-or-i-me-ter}
\hyphenation{de-vices}
\RCS$Revision: 339860 $
\RCS$HeadURL: svn+ssh://alverson@svn.cern.ch/reps/tdr2/papers/TOP-13-001/trunk/TOP-13-001.tex $
\RCS$Id: TOP-13-001.tex 339860 2016-04-23 18:59:54Z alverson $

\providecommand{\Pj}{\HepParticle{j}{}{}}

\newcommand{\AlResultCombined}{\ensuremath{0.26\pm 0.11 }\xspace}
\newcommand{\AlResultCombinedStatSys}{\ensuremath{0.26\pm 0.03\stat \pm 0.10\syst} \xspace}
\newcommand{\AlResultCombinedPvalue}{\ensuremath{4.6\% }\xspace}
\newcommand{\AlResultCombinedSvalue}{\ensuremath{2.0}\xspace}

\newcommand{\AlResultTop}{\ensuremath{0.29\pm 0.11 }\xspace}
\newcommand{\AlResultTopStatSys}{\ensuremath{0.29 \pm 0.03\stat \pm 0.10\syst }\xspace}
\newcommand{\AlResultTopPvalue}{\ensuremath{7.9\% } \xspace}
\newcommand{\AlResultTopSvalue}{\ensuremath{1.8}\xspace}

\newcommand{\AlResultAntiTop}{\ensuremath{0.21\pm 0.14 }\xspace}
\newcommand{\AlResultAntiTopStatSys}{\ensuremath{0.21\pm 0.05\stat \pm 0.13\syst }\xspace}
\newcommand{\AlResultAntiTopPvalue}{\ensuremath{5.0\% }\xspace}
\newcommand{\AlResultAntiTopSvalue}{\ensuremath{2.0}\xspace}
  \newcommand{\Amu}{\ensuremath{A_{\PGm}}\xspace}
  \newcommand{\AmuT}{\ensuremath{A_{\PGm}(\PQt)}\xspace}
  \newcommand{\AmuTbar}{\ensuremath{A_{\PGm}(\PAQt)}\xspace}
  \newcommand{\AmuTplusTbar}{\ensuremath{A_{\PGm}(\PQt+\PAQt)}\xspace}

  \newcommand{\mt}{\ensuremath{m_{\PQb \PGm\PGn}}\xspace}
  \newcommand{\mlbnu}{\mt}
  \newcommand{\mlvb}{\mt}
  \newcommand{\etalvb}{\ensuremath{|\eta_{\PQb \PGm\PGn}|}\xspace}
  \newcommand{\cosThetaPol}{\ensuremath{\cos{\theta^*_{\PGm}}}\xspace}
  \newcommand{\costheta}{\cosThetaPol}
  \newcommand{\cosTheta}{\cosThetaPol}

  \newcommand{\thetaL}{$\theta^*_{\PGm}$}
  \newcommand{\etalj}{\ensuremath{|\eta_{\Pj^\prime}|}\xspace}
  \newcommand{\etabj}{\ensuremath{|\eta_{\PQb}|}\xspace}
  \newcommand{\vtb}{$V_{\mathrm{tb}}$}
  \newcommand{\vtbAbs}{$|V_{\mathrm{tb}}|$}

  \newcommand{\pvmiss}{\ensuremath{\vec{p_\mathrm{T}}\hspace{-1.02em}/\kern 0.5em}\xspace}
  \newcommand{\met}{\ETslash}
  \renewcommand{\MET}{\ETslash}
  \newcommand{\boldMET}{\boldsymbol{\ensuremath{E_{\mathrm{T}}\hspace{-0.9em}/\kern0.55em}\xspace}}

  \newcommand{\PFrelIso}{\ensuremath{I_{\mathrm{rel}}^{\beta-corr.}}\xspace}
  \newcommand{\PFrelIsoRho}{\ensuremath{I_{\mathrm{rel}}^{\rho-corr.}}\xspace}
  \newcommand{\pfPhotonIso}{\ensuremath{I^{\gamma}}\xspace}
  \newcommand{\pfChargedHadronIso}{\ensuremath{I^{\mathrm{ch.\,h}}}\xspace}
  \newcommand{\pfNeutralHadronIso}{\ensuremath{I^{\mathrm{n.\,h}}}\xspace}
  \newcommand{\pfPU}{\ensuremath{I^{\mathrm{PU}}\xspace}}
  \newcommand{\sumPUPT}{\ensuremath{\sum p_T^{PU}}\xspace}
  \newcommand{\sumPUPt}{\sumPUPT}
  \newcommand{\rhoEnergy}{\ensuremath{\rho \times A}}
  \newcommand{\MTW}{\ensuremath{m_{\mathrm{T}}(\PW)}\xspace}
  \newcommand{\mTW}{\MTW}
  \newcommand{\mW}{\ensuremath{m_{\PW}}\xspace}
  \newcommand{\mT}{\ensuremath{m_{\mathrm{T}}}\xspace}
  \newcommand{\pTlepton}{\ensuremath{p_{\mathrm{T}}^{\PGm}}\xspace}
  \newcommand{\pTlj}{\ensuremath{p_{\mathrm{T}}^{\mathrm{j^{\prime}}}}\xspace}
  \newcommand{\pTbj}{\ensuremath{p_{\mathrm{T}}^{\mathrm{b}}}\xspace}
  \newcommand{\pTW}{\ensuremath{p_{\mathrm{T}}^{\mathrm{W}}}\xspace}
  \newcommand{\pTtop}{\ensuremath{p_{\mathrm{T}}^{\mathrm{t}}}\xspace}
  \newcommand{\bdtsigbg}{\ensuremath{\text{BDT}_{\PW/\ttbar}}\xspace}
  \newcommand{\bdtqcd}{\ensuremath{\text{BDT}_{\mathrm{multijet}}}\xspace}

  \newcommand{\qcd}{\ensuremath{QCD}\xspace}
  \newcommand{\QCD}{\qcd}
  \newcommand{\wjets}{\ensuremath{\PW\mathrm{\mbox{+}jets}}\xspace}
  \newcommand{\WJets}{\wjets}
  \newcommand{\zjets}{\ensuremath{\PZ/\PGg^*\mathrm{\mbox{+}jets}}\xspace}
  \newcommand{\ZJets}{\zjets}
  \renewcommand{\ttbar}{\ensuremath{\PQt\PAQt}\xspace}
  \renewcommand{\tt}{\ttbar}
	
  \newcommand{\topfit}{\mbox{\textsc{TopFit}}}
  \newcommand{\Nvertices}{\ensuremath{\mathrm{N_{QCD}}}\xspace}
  \newcommand{\pT}{\pt}

  \newcommand{\ratio}{1.14\xspace}
  \newcommand{\ratiostat}{0.12\xspace}
  \newcommand{\ratiosyst}{0.12\xspace}

  \newcommand{\etaCutST}{2.5\xspace}

  \newcommand{\lumiMuFB}{\ensuremath{19.7 \pm 0.5}\xspace}
  \newcommand{\lumiunc}{2.5\xspace}
  \newcommand{\runrange}{190456-208686\xspace}
  \newcommand{\runRangeFull}{\runrange}

  \newcommand{\qcdresultsmuwsample}{2204.3\xspace}
  \newcommand{\qcdresultsmuwsampleerror}{29.9\xspace}

  \newcommand{\qcdresultselewsample}{206.7\xspace}
  \newcommand{\qcdresultselewsampleerror}{5.0\xspace}

  \newcommand{\qcdresultsmusampleb}{408.294\xspace}
  \newcommand{\qcdresultsmusampleberror}{15.9368\xspace}

  \newcommand{\qcdresultselesampleb}{127.875\xspace}
  \newcommand{\qcdresultselesampleberror}{13.0471\xspace}

  \newcommand{\qcdresultsmusr}{117.892\xspace}
  \newcommand{\qcdresultsmusrerror}{4.22283\xspace}

  \newcommand{\qcdresultselesr}{109.115\xspace}
  \newcommand{\qcdresultselesrerror}{5.0711\xspace}

\newcommand{\AMCATNLO} {a\textsc{mc@nlo}\xspace}

\newcolumntype{d}[1]{D{,}{\,\pm\,}{#1,4} }

\include{numbers}

\newlength\cmsFigWidth
\ifthenelse{\boolean{cms@external}}{\setlength\cmsFigWidth{0.85\columnwidth}}{\setlength\cmsFigWidth{0.4\textwidth}}
\ifthenelse{\boolean{cms@external}}{\providecommand{\cmsLeft}{top}}{\providecommand{\cmsLeft}{left}}
\ifthenelse{\boolean{cms@external}}{\providecommand{\cmsRight}{bottom}}{\providecommand{\cmsRight}{right}}
\cmsNoteHeader{TOP-13-001}
\title{Measurement of top quark polarisation in \texorpdfstring{$t$-channel}{t-channel} single top quark production}

\date{\today}

\abstract{A first measurement of the top quark spin asymmetry, sensitive to the top quark polarisation, in $t$-channel single top quark production is presented. It is based on a sample of pp collisions at a centre-of-mass energy of 8 TeV corresponding to an integrated luminosity of 19.7\fbinv. A high-purity sample of $t$-channel single top quark events with an isolated muon is selected. Signal and background components are estimated using a fit to data. A differential cross section measurement, corrected for detector effects, of an angular observable sensitive to the top quark polarisation is performed. The differential distribution is used to extract a top quark spin asymmetry of $0.26 \pm 0.03\stat \pm 0.10\syst $, which is compatible with a $p$-value of $4.6\%$ with the standard model prediction of $0.44$.
}

\hypersetup{%
pdfauthor={CMS Collaboration},%
pdftitle={Measurement of top quark polarisation in t-channel single top quark production},%
pdfsubject={CMS},%
pdfkeywords={CMS, physics, top physics, polarization}}

\maketitle

\section{Introduction}
\label{sec:introduction}
The top quark is the heaviest elementary particle discovered so far.
Its lifetime (${\approx}4\times 10^{-25}\unit{s}$) is much shorter than the typical timescales of quantum chromodynamics (QCD).
It is therefore the only quark that decays through electroweak interactions before hadronising.
Furthermore, the parity-violating nature of the V--A electroweak interaction at the $\PW\PQt\PQb$ vertex means that only left-handed quarks are expected at this vertex.
 Thus, top quark decay products retain memory of the top quark spin orientation in their angular distributions. This fact turns the top quark into a powerful probe of the structure of the electroweak $\PW\PQt\PQb$ vertex.

\begin{figure}[h]
	    \begin{center}
	    \includegraphics[width=0.30\textwidth]{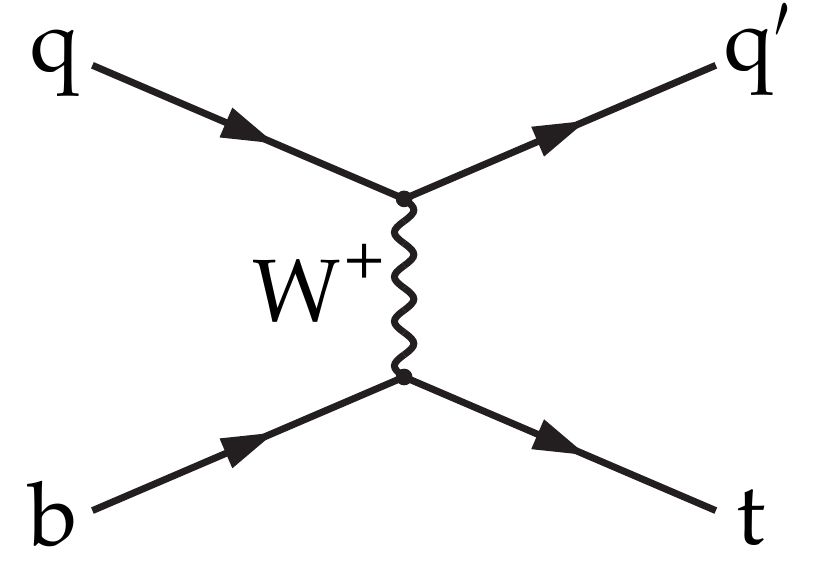}
	    \quad \quad \quad \quad
	    \includegraphics[width=0.30\textwidth]{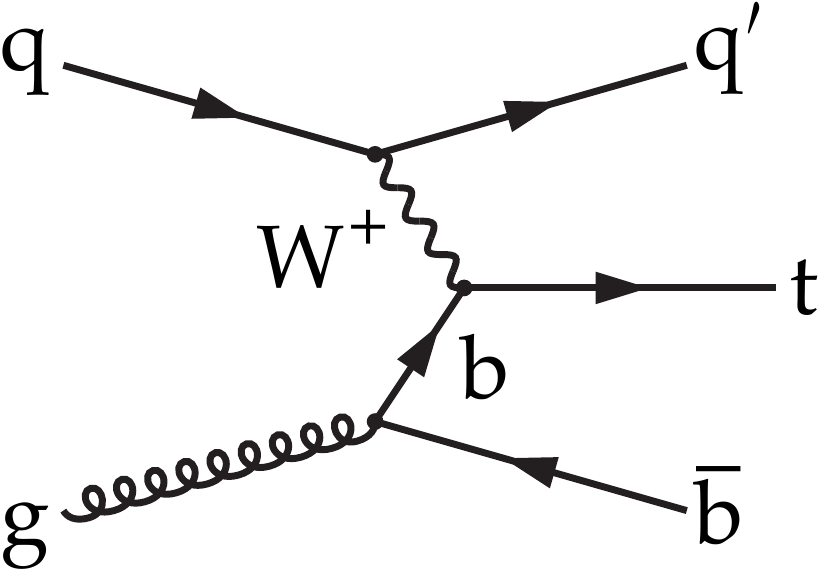}
	    \caption{\label{fig:FG} Feynman diagrams for single top quark production in the $t$-channel: (left)~(2)$\rightarrow$(2) and (right)~(2)$\rightarrow$(3) processes. Similar diagrams are expected for top antiquark production.}
	  \end{center}
	\end{figure}

In electroweak $t$-channel single top quark production, shown in Fig.~\ref{fig:FG}, the standard model (SM) predicts that produced top quarks are highly polarised, as a consequence of
 the V--A coupling structure, along the direction of the momentum of the spectator quark ($\PQq^{\prime}$),
 which recoils against the top quark~\cite{Mahlon:1999gz,Jezabek:1994zv}.
 However, new physics models could also lead to a depolarisation in production by altering the coupling structure~\cite{AguilarSaavedra:2010nx,AguilarSaavedra:2008gt,AguilarSaavedra:2008zc,Bach:2012fb}.

In this analysis, the top quark spin asymmetry

\begin{equation}
A_{X} \equiv\frac{1}{2}\, P_{\PQt}\, \alpha_{X} = \frac{N(\uparrow)-N(\downarrow)}{N(\uparrow)+N(\downarrow)} \,
\label{eq:top-quark-asymmetry}
\end{equation}
is used to probe the coupling structure, where $P_{\PQt}$ represents the top quark polarisation in production and
$\alpha_{X}$ denotes the degree of the angular correlations of one of its decay products, denoted $X$ (where for this analysis $X=\Pgm$),
 with respect to the spin of the top quark, the so-called spin-analysing power. The variables ${N(\uparrow)}$ and ${N(\downarrow)}$ are defined, for each top quark decay product from the decay chain $\PQt \rightarrow \PQb \PW \rightarrow \PQb \Pgm \Pgn$, as the number of instances in which that decay product is aligned or antialigned, respectively, relative to the direction of the recoiling spectator quark momentum.

In this analysis, the muon is chosen as the top quark spin analyser because leptons have the highest spin-analysing power and since the muon identification efficiency is very high in the CMS detector. The spin-analysing power is exactly 1 at leading order~(LO) in the SM. Its value can be modified by new physics that may be characterised by anomalous top quark coupling models arising from an effective extension of the coupling structure of the $\PW\PQt\PQb$ vertex~\cite{AguilarSaavedra:2008zc}.

The measurement of the top quark spin asymmetry, measured in $t$-channel single top quark events with one isolated muon in the final state, is the subject of this paper. The asymmetry is measured for top quark and antiquark events separately to be sensitive to potential CP-violation, which is predicted in some new physics models.

The analysis strategy is as follows: after applying an event selection designed to obtain a set of relatively high purity $t$-channel single top quark events, the signal and background composition of data is estimated using a binned likelihood fit. A top quark candidate is then reconstructed and the angle between the muon and the recoiling jet calculated in the top quark rest frame.

An unfolding technique is applied to obtain a differential cross section measurement of this angular distribution at parton level. From the unfolded distribution, the top quark spin asymmetry, which is directly related to the polarisation through Eq.~(\ref{eq:top-quark-asymmetry}), is calculated for top quark and antiquark events, and their combination.

\section{The CMS detector and event reconstruction}

The central feature of the CMS apparatus is a superconducting solenoid of 6\unit{m} internal diameter, providing a magnetic field of 3.8\unit{T}. Within the solenoid volume are a silicon pixel and strip tracker, a lead tungstate crystal electromagnetic calorimeter (ECAL), and a brass and scintillator hadron calorimeter (HCAL), each composed of a barrel and two endcap sections. Muons are measured in gas-ionisation detectors embedded in the steel flux-return yoke outside the solenoid. Extensive forward calorimetry complements the coverage provided by the barrel and endcap detectors. A more detailed description of the CMS detector, together with a definition of the coordinate system used and the relevant kinematic variables, can be found in Ref.~\cite{Chatrchyan:2008zzk}.

The particle-flow algorithm~\cite{CMS-PAS-PFT-09-001,CMS-PAS-PFT-10-001} reconstructs and identifies each individual particle in an event  with an optimised combination of information from the various elements of the CMS detector. The energy of photons is directly obtained from the ECAL measurement. The energy of electrons is determined from a combination of the electron momentum at the primary interaction vertex, as determined by the tracker, the energy of the corresponding ECAL cluster, and the energy sum of all bremsstrahlung photons spatially compatible with originating from the electron track. The energy of muons is obtained from the curvature of the corresponding track. The energy of charged hadrons is determined from a combination of their momentum measured in the tracker and the matching ECAL and HCAL energy deposits, corrected for the response function of the calorimeters to hadronic showers. Finally, the energy of neutral hadrons is obtained from the corresponding corrected ECAL and HCAL energy. To mitigate the effect of pileup, i.e. additional proton-proton collisions whose signals in the detector sum to the products of the primary interaction that triggered the event, charged particles associated to non-leading primary vertices are vetoed.

The missing transverse momentum vector, \pvmiss, is defined as the projection onto the plane perpendicular to the beams of the negative vector sum of the momenta of all reconstructed particles in an event. Its magnitude is referred to as missing transverse energy (\met).

\section{Data and simulated samples}

This study is based on the proton-proton collision data set recorded by the CMS detector at the CERN LHC in 2012 at a centre-of-mass energy 8\TeV, corresponding to an integrated luminosity of \lumiMuFB\fbinv~\cite{lumi}.

Single top quark $t$-channel events from Monte Carlo (MC) simulation are generated with the next-to-leading-order (NLO) MC
generator \POWHEG~1.0~\cite{Alioli:2010xd,Alioli:2009je,Frixione:2007vw}, interfaced with \PYTHIA~6.4~\cite{Sjostrand:2006za} for the parton showering, in which $\Pgt$ lepton decays are modelled with \TAUOLA~\cite{Jadach:1990mz}.
The 5-flavour scheme (5FS) is used in the generation, i.e. inherent \PQb quarks are considered among the incoming particles as in Fig.~\ref{fig:FG}~(left).
 As an alternative NLO generator, used to assess the dependence of the analysis on the modelling of signal,
 we use \AMCATNLO~2.1.2~\cite{Alwall:2014hca} interfaced with \PYTHIA~8.180~\cite{Sjostrand:2014zea},
 with the 4-flavour scheme (4FS), i.e. \PQb quarks in the initial state are only produced via gluon splitting as in Fig.~\ref{fig:FG}~(right).
 The measured results are compared with predictions from the aforementioned NLO generators and the LO generator \COMPHEP~4.5~\cite{Boos:2004kh},
 interfaced with \PYTHIA~6, with a matching procedure between LO 5FS and 4FS diagrams based on the transverse momentum \pt\ of the associated \PQb quark~\cite{boos2000}.
 Special samples are generated using \COMPHEP~4.5 including a $\PW\PQt\PQb$ coupling with anomalous structure.

 Several SM processes are taken into account as backgrounds in the analysis. The \POWHEG~1.0 generator interfaced with \PYTHIA~6 is also used to
model the \PW-associated ($\PQt\PW$) and $s$-channel single top quark background events.
 The \ttbar, \PW boson in association with jets (\wjets), and Drell--Yan in association with jets (\zjets) processes
 are generated with \MADGRAPH~5.1~\cite{madgraph} interfaced with \PYTHIA~6. \TAUOLA is used to simulate $\PGt$ lepton decays.
Up to three (four) additional partons are generated at matrix-element (ME) level in \ttbar\ (\wjets\ and \zjets) events.
A procedure, implemented during event generation, based on the so-called ``MLM prescription''~\cite{mlm,Alwall:2007fs},
avoids double counting jets generated simultaneously by the ME and by the parton shower (PS) simulations.
 An alternative sample of \wjets\ generated with \SHERPA~1.4.0 at NLO~\cite{Hoche:2010pf,Hoeche:2012ft} is used to compare the modelling of this background.
 Diboson production ($\PW\PW$, $\PW\PZ$, $\PZ\PZ$) is simulated using \PYTHIA~6.
Multijet events (i.e. events with the muon not originating from a leptonically decaying $\PW$ or $\PZ$ boson) are modelled using statistically independent samples in data, as detailed in Section~\ref{sec:qcd}.
Other special samples of signal and background are generated with different values for generator parameters (e.g. top quark mass, renormalisation and factorisation scales, etc.), and used to estimate the corresponding systematic uncertainties.

{\tolerance=1200
All single top quark processes are normalised to approximate next-to-next-to-leading-order (NNLO) predictions~\cite{Kidonakis:2012db}
($\sigma_{{\operatorname{\mathit{t}\mbox{-}\mathrm{channel}}}}=87.1\unit{pb}$, $\sigma_{\operatorname{\mathit{s}\mbox{-}\mathrm{channel}}}=5.55\unit{pb}$, $\sigma_{{\rm tW}}=22.2\unit{pb}$). Top quark pair production is normalised to
 a complete NNLO prediction in QCD that includes soft gluon resummation to next-to-next-to-leading-log order, as calculated
 with the {\sc Top++2.0} program~\cite{Czakon:2011xx} ($\sigma_{\ttbar}=252.9\unit{pb}$).
 The \wjets\ and \zjets\ production cross sections times branching fraction are calculated at NNLO with {\sc fewz}~\cite{Gavin:2010az}
 ($\sigma_{\wjets}\, \mathcal{B}(\PW \to \ell \nu)=37\,509\unit{pb}$, and $\sigma_{\zjets}\, \mathcal{B}(\PZ/\PGg^*\to \ell^{+} \ell^{-})=3504\unit{pb}$ at a generator-level threshold of $m_{\ell^{+} \ell^{-}}>50\GeV$,
 where $\ell = \Pe$, $\PGm$, or $\PGt$).
 The diboson cross sections are calculated at NLO with \MCFM~5.8~\cite{Campbell:2010ff} ($\sigma_{\PW\PW}=54.8\unit{pb}$, $\sigma_{\PW\PZ}=33.2\unit{pb}$, and $\sigma_{\PZ\PZ}=8.1\unit{pb}$).
\par}

The effect of pileup is evaluated using a simulated sample of minimum-bias events produced using \PYTHIA~6,
superimposed onto the events in the simulated samples described above, taking into account in-time and out-of-time pileup contributions.
The events are then reweighted to reproduce the true pileup distribution inferred from the data.
 The procedure is validated by comparing the number of observed primary vertices between data and simulation.

All generated events undergo a full \GEANTfour~\cite{geant} simulation of the detector response.

\section{Event selection}
\label{sec:selection}
The study presented here focuses on the $ \PQt \to \PQb \PW \to \PQb \PGm \PGn$ decay
channel. Signal events are characterised by exactly one isolated muon, large $\MET$ (originating from the neutrino in the leptonic decay of the W boson),
one central b jet from the top quark decay, and an additional untagged jet ($\Pj^{\prime}$) from the spectator quark ($\PQq^{\prime}$)
from the hard-scattering process, which is preferentially produced in the forward region of the detector.
 A second b jet produced in association with the top quark can also be present in the detector, although it yields a softer $\pt$ spectrum
relative to the b jet from the top quark decay.
 The event selection applied in the measurement of the production cross section in the same channel~\cite{Chatrchyan:2012ep} is closely followed.

Trigger selection is based on the presence of at least one isolated muon with
$\pt> 24\GeV$ and $|\eta| < 2.1$.

One isolated muon candidate is required to originate from the leading primary vertex, which is defined as the vertex with the largest value of the summed $\pT^{2}$ of its associated charged tracks.
 Muon candidates are accepted if they pass the following requirements:
 $\pt$ of at least 26\GeV, $|\eta| < 2.1$, quality and identification criteria optimised for the selection of prompt muons, and a relative isolation requirement of $I_\mathrm{rel} <0.12$.
 The relative isolation, $I_\mathrm{rel}$, is defined by the scalar sum, divided by the $\pT$ of the muon, of the transverse energies deposited by stable charged hadrons, photons,
 and neutral hadrons where deposits linked to pileup are subtracted within a cone of radius
 $\Delta R = \sqrt{\smash[b]{(\Delta\eta)^2+(\Delta\phi)^2}} = 0.4$ (where $\phi$ is the azimuthal angle in radians) around the muon direction.
 Events are rejected if an additional muon or electron candidate is present.
 The selection requirements for these additional electrons/muons are as follows:
 looser identification and isolation criteria, $\pt >$ 10~(20)\GeV for muons (electrons), and $|\eta| < 2.5$.

Jets are reconstructed from the particle-flow candidates and clustered with the anti-$\kt$ algorithm~\cite{Cacciari:2008gp,Cacciari:2011ma} with a distance parameter of 0.5.
The influence of pileup is mitigated using the charged hadron subtraction technique~\cite{CMS-PAS-JME-14-001}. The jet momentum is determined as the vectorial sum of all particle momenta in the jet.
An offset correction is applied to the transverse jet momenta to account for contributions from pileup.
Further corrections are applied to account for the non-flat detector response in $\eta$ and $\pT$ of the jets.
The corrected jet momentum is found from simulation to be within 5\% to 10\% of the true momentum over the whole \pt spectrum and detector acceptance.
The corrections are propagated to the measured \pvmiss as it depends on the corrected jets through the clustered tracks.
 Additional selection criteria are applied to each event to remove spurious jet-like features originating from isolated noise patterns in certain HCAL regions.
 The analysis considers jets within $|\eta|<4.5$ whose calibrated transverse energy is greater than 40\GeV.
 The event is accepted for further analysis only if at least two such jets are present.

To reduce the large background from \wjets events, a \PQb tagging algorithm based on combined information
from secondary vertices and track-based lifetimes~\cite{Chatrchyan:2012jua,CMS-PAS-BTV-13-001} is used.
 A tight selection is applied on the b tagging discriminant, which corresponds to an efficiency of ${\approx} 50\%$ for jets originating from true b quarks and a mistagging rate of ${\approx} 0.1\%$ for other jets in the signal simulation.
 The \PQb tagging performance in simulation is corrected to better match the performance observed in data~\cite{CMS-PAS-BTV-13-001},
using scale factors that depend on the $\pt$ and $\eta$ of the selected jets.

Corrections are applied to the simulation, where necessary, to account for known differences relative to data.
Single-muon trigger efficiencies and lepton reconstruction and identification efficiencies are estimated with a ``tag-and-probe'' method~\cite{Chatrchyan:2012xi,CMS-DP-2013-009} from \ZJets data.
 B tagging and misidentification efficiencies are estimated by dedicated analyses performed with statistically independent selections~\cite{CMS-PAS-BTV-13-001}.
 A smearing of the jet momenta is applied to account for the known difference in jet energy resolution (JER) in simulation compared to data~\cite{Chatrchyan:2011ds}.
 The effects of all these corrections are found to be small.

To classify signal and control regions, different event categories, denoted ``$N$jets $M$tag(s)'' are defined, where $N$ is the total number of selected jets (2 or 3)
and $M$ is the number of those jets passing additionally the \PQb tagging requirements (0, 1, or 2). The ``2jets 1tag'' category defines the region used for signal extraction,
whereas the other categories, enriched in background processes with different compositions,
 are used for the control samples discussed in Section~\ref{sec:control}.
 The ``2jets 1tag'' category is separated into a control region and a signal region, depending on the value of a multivariate discriminant, described below.

In the ``2jets 1tag'' category, a top quark candidate is reconstructed from the \PQb jet, the muon, and a neutrino candidate.
A neutrino candidate is constructed as described in Ref.~\cite{Chatrchyan:2011vp}. The neutrino $p_{z}^{\nu}$ momentum is found by requiring a \PW\ boson mass constraint from momentum conservation using the muon and missing transverse momenta.
In the other categories, the jet with the highest value of the \PQb tagging discriminant is used for top quark reconstruction.

Multijet events are suppressed by setting a threshold on the output of a dedicated boosted decision tree (\bdtqcd), trained using the following input variables:
\begin{itemize}
\item the missing transverse energy, \MET;
\item the invariant mass of the top quark candidate, \mlvb;
\item the transverse mass of the \PW\ boson candidate, \\
$\mTW = \sqrt{ \smash[b]{ \left(p_{\mathrm{T}}^{\mu} + \MET \right)^2 - \left( p_x^{\mu} + \pvmiss {}_{,x} \right)^2 - \left( p_y^{\mu} + \pvmiss {}_{,y} \right)^2 } }$;
\item the transverse momentum of the untagged jet, \pTlj;
\item the event isotropy, defined as $({\cal S}_{\mathrm{max}}-{\cal S}_{\mathrm{min}})/{\cal S}_{\mathrm{max}}$ with ${ {\cal S} \equiv \displaystyle\sum_{i  }^{\PGm,\, \mathrm{jets}} | \vec{n} \cdot \vec{p}_i | }$, where the unit vector in the transverse $r$--$\phi$ plane, $\vec{n}=(\cos\phi,\sin\phi)$, can be chosen to either maximise or minimise $\cal S$.
\end{itemize}

To reject background events, a second boosted decision tree, \bdtsigbg, is used to separate signal from \ttbar and \wjets events.
Training is performed with the following input observables:
\begin{itemize}
\item the invariant mass of the top quark candidate, \mlvb;
\item the absolute pseudorapidity of the untagged jet, \etalj;
\item the absolute pseudorapidity of the b-tagged jet, \etabj;
\item the invariant mass of the $\PQb$-tagged jet from the summed momenta of the clustered tracks, $m_{\PQb}$;
\item the transverse momentum of the muon, \pTlepton;
\item the transverse momentum of the b-tagged jet, \pTbj;
\item the transverse mass of the \PW\ boson candidate, $\mTW$;
\item the missing transverse energy, \MET;
\item the total invariant mass of the top quark candidate and the untagged jet system, $\hat{s}$;
\item the transverse momentum of the hadronic final-state system, $H_\mathrm{T} = (\vec{p}_{\PQb} + \vec{p}_{\Pj^{\prime}})_{\mathrm{T}}$.
\end{itemize}

\begin{figure}[hbtp]
\begin{center}
\includegraphics[width=.49\textwidth]{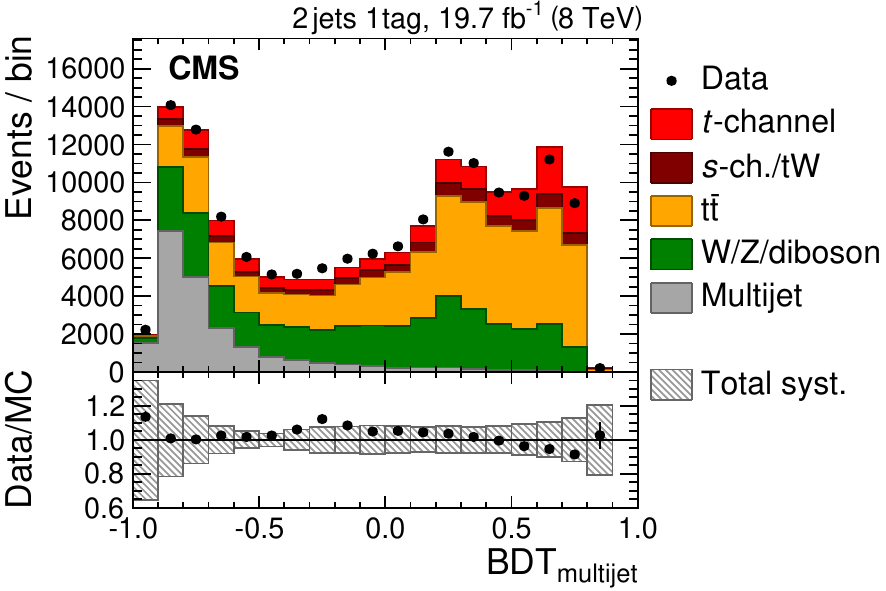}
\includegraphics[width=.49\textwidth]{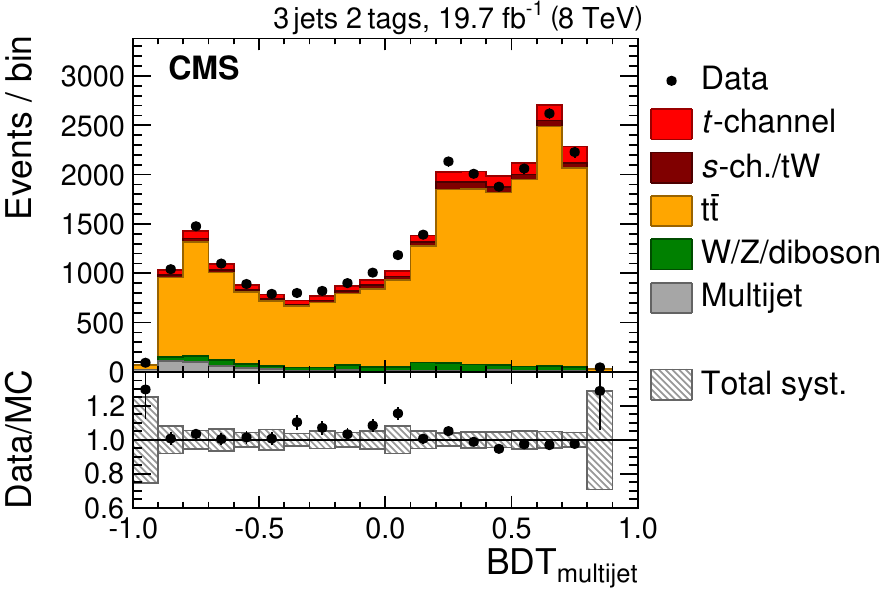}
\caption{Distributions of the \bdtqcd discriminant in the (left)~``2jets 1tag'' and (right)~``3jets 2tags'' categories.
The predictions are normalised to the results of the fit described in Section~\ref{sec:fit}.
The bottom panels in both plots show the ratio between observed and predicted event counts,
with a shaded area to indicate the systematic uncertainties affecting the background prediction and vertical bars indicating statistical uncertainties.}
\label{fig:bdt_qcd_output_control}
\end{center}
\end{figure}

\begin{figure}[htp]
\begin{center}
\includegraphics[width=0.49\textwidth]{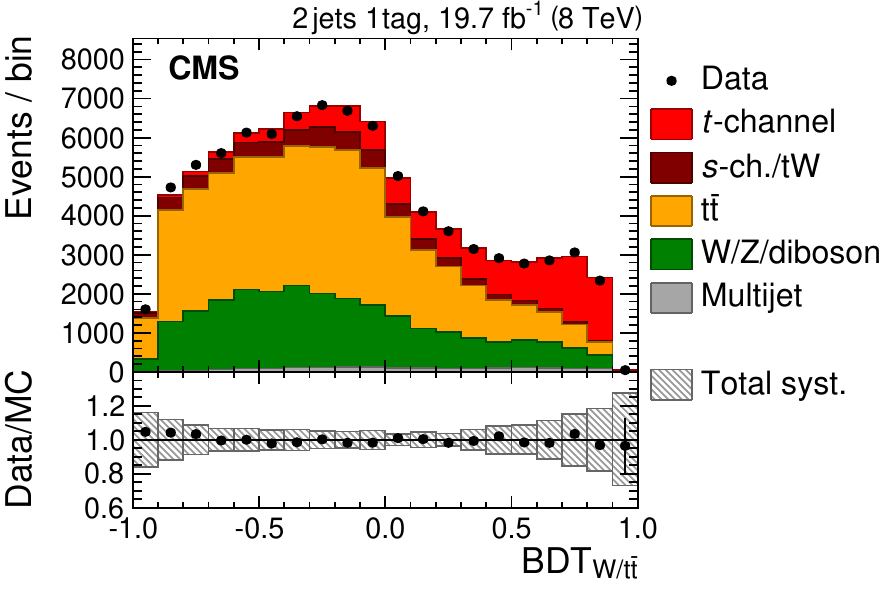}
\includegraphics[width=0.49\textwidth]{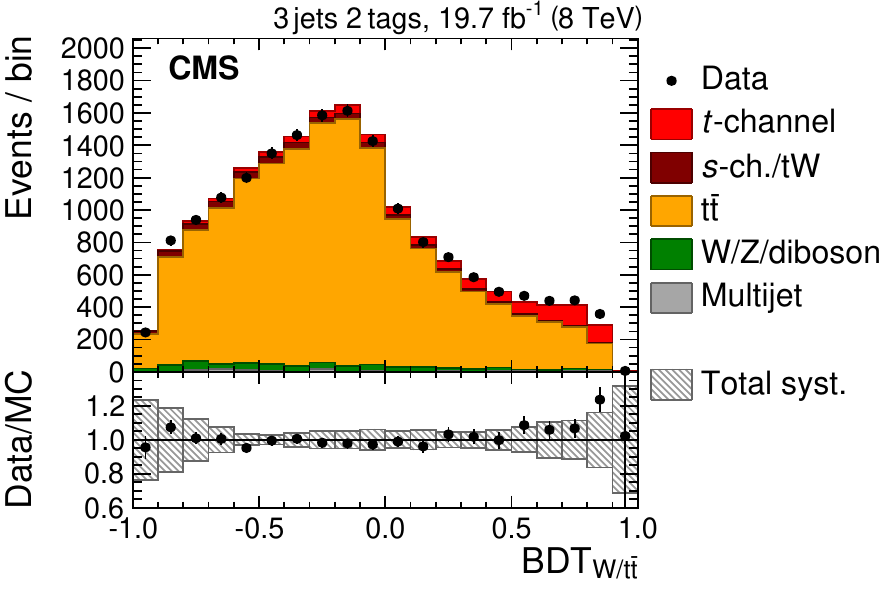}
\caption{\label{fig:BDT_sig}Distributions of the \bdtsigbg discriminant in the (left)~``2jets 1tag'' and (right)~``3jets 2tags'' categories.
The predictions are normalised to the results of the fit described in Section~\ref{sec:fit}.
The bottom panels in both plots show the ratio between observed and predicted event counts,
with a shaded area to indicate the systematic uncertainties affecting the background prediction and vertical bars indicating statistical uncertainties.}
\end{center}
\end{figure}

By construction, the BDT discriminant ranges between $+1$ and $-1$,
with the algorithm trained such that the resulting distribution peaks at a high BDT discriminant value for signal-like events and at a low value for background-like events.
The distribution of the \bdtqcd discriminant is shown in Fig.~\ref{fig:bdt_qcd_output_control} in two categories,
with the multijet events shape and normalisation extracted as described in Section~\ref{sec:qcd}.
To reject multijet events, we only use events that pass the threshold $\bdtqcd~\mathrm{discriminant} > -0.15$ in the analysis.
Figure~\ref{fig:BDT_sig} shows the distribution of the \bdtsigbg discriminant in the ``2jets 1tag'' and ``3jets 2tags'' categories
after applying the selection requirement on the $\bdtqcd$ discriminant.

\begin{figure}[hbtp]
\begin{center}
\includegraphics[width=0.49\textwidth]{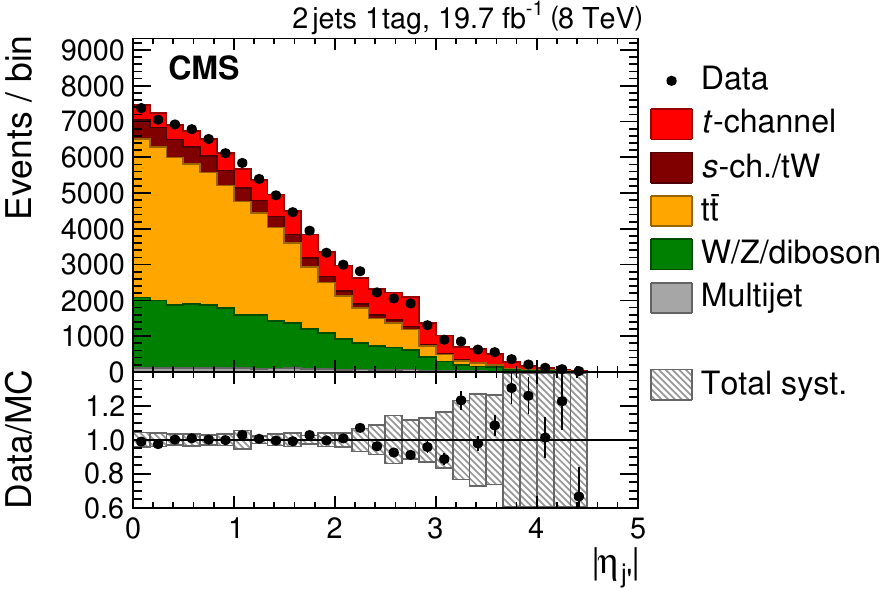}
\includegraphics[width=0.49\textwidth]{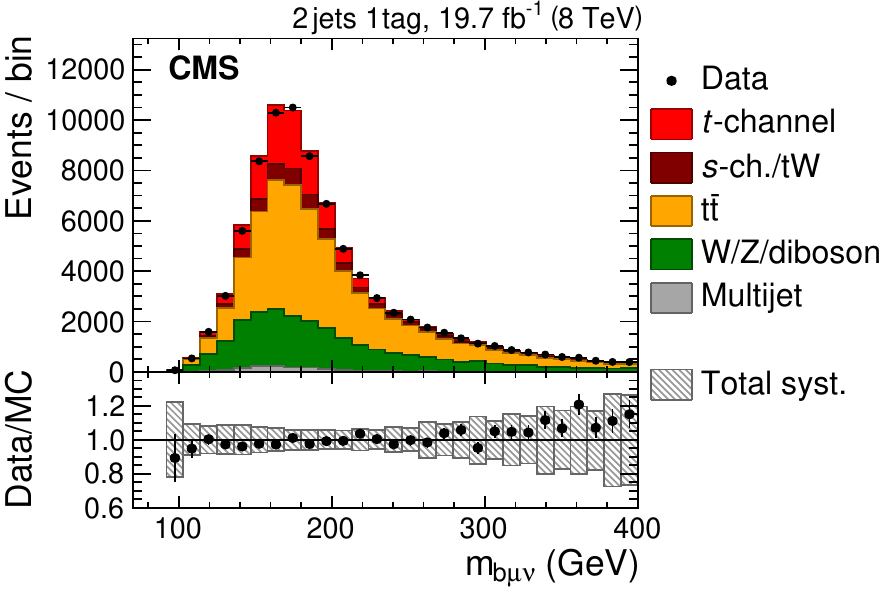}
\caption{Distributions of the \etalj (left) and \mlvb (right) variables in the ``2jets 1tag'' category.
In both plots, the rejection of multijet events is performed by requiring $\bdtqcd > -0.15$.
The predictions are normalised to the results of the fit described in Section~\ref{sec:fit}.
The bottom panels in both plots show the ratio between observed and predicted event counts,
with a shaded area to indicate the systematic uncertainties affecting the background prediction and vertical bars indicating statistical uncertainties.}
\label{fig:eta_topmass}
\end{center}
\end{figure}

Figure~\ref{fig:eta_topmass} shows the distributions of the \etalj and \mlvb variables in the ``2jets 1tag'' category.
These variables have the highest ranking in the decision of the \bdtsigbg.

To select a signal-enhanced phase space, an additional selection is imposed on the \bdtsigbg discriminant.
The optimal working point is found to be $\bdtsigbg~\mathrm{discriminant}>0.45$ by studying the analysis sensitivity with pseudo-data from simulated events.

All BDT input variables are found to be well modelled by the MC simulation.
The BDTs are trained and tested on statistically independent samples, with no overtraining observed.

\section{The \texorpdfstring{$\cosThetaPol$}{cos(theta-pol)} distribution of top quark decay products}
\label{sec:polarization}
The angle between a top quark decay product $X$
($\PW$, $\ell$, $\PGn$,  or $\PQb$) and an arbitrary polarisation axis $\vec{s}$ in the top quark rest frame, $\theta^{*}_X$,
is distributed according to the following differential cross section:

\begin{equation}
\frac{1}{\sigma}\frac{\rd\sigma}{\rd\cos\theta^{*}_X} =
\frac{1}{2}(1+P^{(\vec{s})}_{\PQt}\alpha_X\cos\theta^{*}_X) =
\left(\frac{1}{2}+A_X \cos\theta^{*}_X\right) .
\label{eq:cosThetaDistr}
\end{equation}

The variable $P^{(\vec{s})}_{\PQt}$ denotes the single top quark polarisation along the chosen axis, and
$\alpha_X$ the spin-analysing power as defined in Section~\ref{sec:introduction}.
In the SM, the top quark
spin tends to be aligned with the direction of the spectator quark momentum, resulting in a high degree of polarisation.
 Hence, an excess of events where the spectator quark momentum is
antialigned with the top quark spin would clearly indicate an anomalous coupling structure.
 Single top quark polarisation is studied in the $t$-channel
process through the angular asymmetry \Amu of the muon,
 with the polarisation axis defined as pointing along the untagged jet ($\Pj^{\prime}$) direction in the top quark rest frame.

\begin{figure}[hbtp]
  \begin{center}
\includegraphics[width=.49\textwidth]{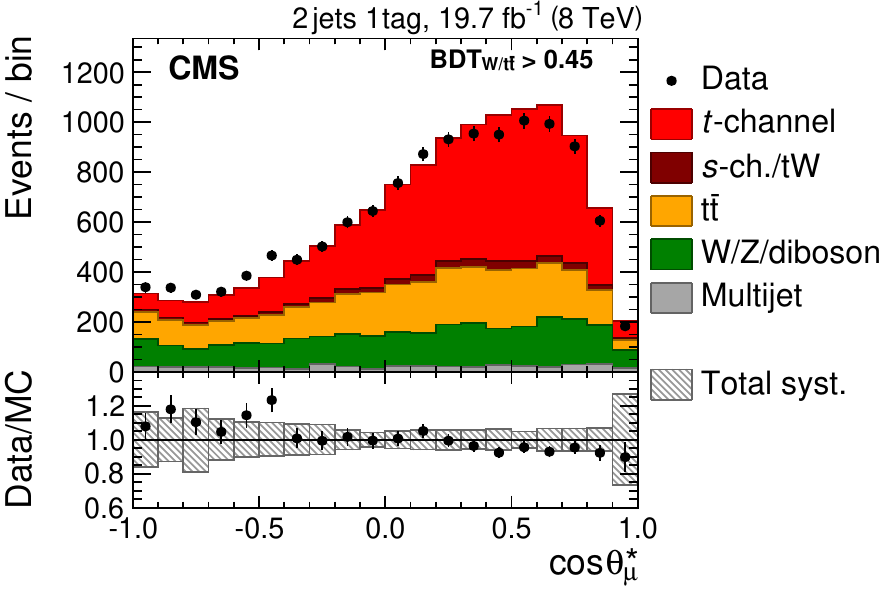}
\includegraphics[width=.49\textwidth]{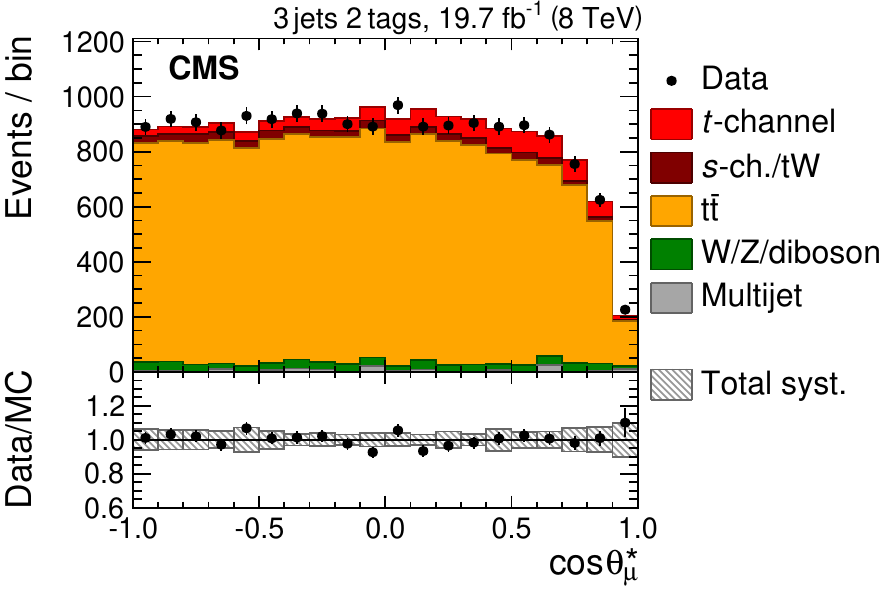}
\caption{Distributions of \cosThetaPol\ in the (left)~signal region defined by $\bdtsigbg>0.45$ in the ``2jets 1tag'' category
and (right)~``3jets 2tags'' control region. In both plots, the rejection of multijet events is performed by requiring $\bdtqcd > -0.15$.
The predictions are normalised to the results of the fit described in Section~\ref{sec:fit}.
The bottom panels in both plots show the ratio between observed and predicted event counts,
with a shaded area to indicate the systematic uncertainties affecting the background prediction, and vertical bars indicating statistical uncertainties.}
\label{fig:cosTheta}
  \end{center}
\end{figure}

Figure~\ref{fig:cosTheta} shows the reconstructed distribution of \costheta\ in the ``2jets 1tag'' (for $\bdtsigbg>0.45$) and ``3jets 2tags'' categories.
The observed distribution is expected to differ from the parton-level prediction because of detector effects and the kinematic selection applied,
with the most significant effect being the relatively small number of selected events close to $\costheta = 1$.
An overall trend in the ratio between data and simulation is observed that suggests a slightly less asymmetric shape than predicted by the SM.

In this analysis, a $\chi^{2}$-fit is performed of the unfolded \cosThetaPol differential cross section to estimate \Amu based on Eq.~(\ref{eq:cosThetaDistr}).

\section{Studies of background modeling}
\label{sec:control}
Statistically independent control samples are used for several purposes in this analysis.
Samples in which the isolation requirement on the muon is inverted are used
 to extract templates for estimating the contamination by multijet events,
 while samples with different jet and b-tagged jet multiplicities are used to validate the simulation of \wjets\ and \ttbar\ events,
 or to provide additional constraints on the in situ determination of background and signal strengths relative to the SM.

\subsection{Estimation of multijet events background}
\label{sec:qcd}

The yield of the multijet events background in the different categories is measured by
 performing fits to the \bdtqcd\ discriminant distributions for each
 ``$N$jet $M$tag'' category where a significant contamination from this process is expected.
 A binned maximum-likelihood (ML) fit to the data is performed using two components: multijet events (unconstrained)
 and the sum of all other processes (constrained to be within ${\pm}20\%$ of the expected yield using a log-normal prior that the fit constrains further). The latter category includes the signal.
A normalised distribution (template) for the sum of other processes is taken from simulation.
The multijet events template is obtained from a statistically independent multijet-enriched data sample with an inverted isolation requirement, as defined above.
It is verified that the \bdtqcd\ discriminant distributions for multijet events are not significantly affected by this altered event selection.

Uncertainties on the multijet events yields
are estimated conservatively to be ${\pm} 50\%$. In addition, an uncertainty on the shape is taken into account by using a modified inverted isolation requirement.
Together, these are used to estimate the systematic uncertainty associated with this procedure, as discussed in Section~\ref{sec:systematics}.

\subsection{\texorpdfstring{\wjets}{W+jets} model validation and correction}
\label{sec:wjets}

After estimating the multijet events contribution to the signal region, the agreement between the expectations and the data is verified
 in several control regions for all the \bdtsigbg\ inputs, for the \bdtsigbg\ response, for \costheta, and for a number of additional variables.
 Among all the control regions considered, \costheta\ and \pT\ of the reconstructed W~boson are observed to be mismodelled in the ``2jets 0tags'' control region;
 this region is expected to be enriched in \wjets\ events.

A similar disagreement between data and the \MADGRAPH\ prediction in the \costheta\ distribution is observed in data collected at $\sqrt{s}=7\TeV$
in the context of a different analysis~\cite{Chatrchyan:2012ep}.
Investigations using different MC generators and their associated settings show that \SHERPA~\cite{Gleisberg:2008ta}
provides a better description of \costheta\ in this control region at both centre-of-mass energies.

Although this control region is not used in the fit,
additional investigations have been performed to check whether this mismodelling can potentially affect the signal region.
 The \MADGRAPH\ and \SHERPA\ samples are found to differ mostly in the \costheta distribution for events with a $\PW$ boson produced in association with jets
 from gluon fragmentation, which constitutes a major component of the ``2jets 0tags'' region, but is a very small fraction of the ``2jets 1tag'' signal region.

 In the kinematic region studied by this analysis, \MADGRAPH\ reproduces the \wjets\ kinematic distributions better than \SHERPA.
 Moreover, for computational reasons, the approximation of using b and c quarks as massless in the generation of the \SHERPA\ samples
 causes the relative fraction of heavy quarks to be unrealistically large.
 For these reasons, \MADGRAPH\ is chosen as the default generator in this analysis,
 and a reweighting of the \wjets\ events simulated with \MADGRAPH\ is performed in all signal and control regions
 using the event ratio between the two generators as a function of \costheta, separately for each flavour component,
 in the ``2jets 0tags'' control region.

\subsection{\texorpdfstring{$\ttbar$}{t-tbar} model validation}

To validate the modelling of \ttbar\ events, we compare simulated events to data in the ``3jets 2tags'' control region for the most relevant observables.
In particular, Figs.~\ref{fig:bdt_qcd_output_control}~(right),
\ref{fig:BDT_sig}~(right), and~\ref{fig:cosTheta}~(right) show the \bdtqcd\ and \bdtsigbg\ discriminants, and the \costheta\ distribution, respectively.
This control region is also used in the fit described in Section~\ref{sec:fit}.

The \MADGRAPH\ model of \ttbar\ production is known to predict a harder top quark \pT\ (\pTtop) spectrum than observed in data~\cite{Chatrchyan:2012saa,Khachatryan:2015oqa}. The spectrum of generator-level top quarks in \ttbar\ events is therefore reweighted
 so that it reproduces the measured differential cross section as a function of \pTtop.

In conclusion, the \ttbar\ modelling provided by \MADGRAPH\ after applying the \pTtop reweighting is found to be in reasonable agreement with data.

\section{Extraction of signal and background yields}
\label{sec:fit}
The signal and background components are estimated by means of a simultaneous ML fit
to the distribution of the \bdtsigbg discriminant in the ``2jets 1tag'' and ``3jets 2tags'' regions. The inclusion of the \ttbar-dominated ``3jets 2tags'' region in the fit provides an additional constraint on the \ttbar background. This also reduces correlations of the estimated \ttbar yield with other contributions.

For all background processes, except the multijet events background, templates from the MC samples are used.
The multijet events template is obtained from data by inverting the isolation selection,
as discussed previously, and its normalisation is kept fixed to the estimated yield described in Section~\ref{sec:qcd}.
 To reduce the number of free parameters,
 several processes that have a similar distribution in both \costheta\ and the \bdtsigbg discriminant are merged into a single contribution:
\begin{itemize}
\item Signal: $t$-channel single top quark production, treated as unconstrained.
\item Top quark background: \ttbar, $s$-channel and $\PQt\PW$ single top quark production,
with their relative fractions taken from simulation; a constraint of $\pm 20\%$ using a log-normal prior is applied.
\item $\PW$/$\PZ$/diboson: \wjets, \zjets, and diboson production, with their relative fractions taken from simulation, have a constraint of ${\pm} 50\%$ using a log-normal prior.
\end{itemize}

The results of the three fits, and the post-fit uncertainties for top quark events, top antiquark events, and their combination, are presented in Table~\ref{tab:fit_bdt}
as scale factors to be applied to simulation yields, while Table~\ref{tab:yields} shows the number of events exceeding the threshold on the $\bdtsigbg$ discriminant $ {>}0.45$.
The number of top quark events is greater than the number of top antiquark events due to the up-quark density being larger than either the down-quark or up-antiquark densities at large values of Bjorken $x$ in the incoming protons.

\begin{table}
\begin{center}
\topcaption{Estimated scale factors and uncertainties from the simultaneous maximum-likelihood fit to the distribution of the \bdtsigbg discriminant in the ``2jets 1tag`` and ``3jets 2tags`` categories.}
\begin{tabular}{ r|ccc }
Processes & $\mathrm{t}$ & $\mathrm{\bar{t}}$ & $\mathrm{t+\bar{t}}$ \\
\hline
Signal                              & 1.10 $\pm$ 0.03 & 1.20 $\pm$ 0.05 & 1.13 $\pm$ 0.03 \\
Top quark bkg.                            & 1.06 $\pm$ 0.02 & 1.08 $\pm$ 0.02 & 1.07 $\pm$ 0.01 \\
$\PW$/$\PZ$/diboson                                  & 1.26 $\pm$ 0.05 & 1.21 $\pm$ 0.06 & 1.24 $\pm$ 0.04 \\
\end{tabular}
\label{tab:fit_bdt}
\end{center}
\end{table}

\newcolumntype{L}[1]{>{\raggedright\let\newline\\\arraybackslash\hspace{0pt}}m{#1}}
\newcolumntype{C}[1]{>{\centering\let\newline\\\arraybackslash\hspace{0pt}}m{#1}}
\newcolumntype{R}[1]{>{\raggedleft\let\newline\\\arraybackslash\hspace{0pt}}m{#1}}

\begin{table}[h]
\begin{center}
\topcaption{The expected number of signal and background events in the ``2jets 1tag'' signal region ($\bdtsigbg>0.45$) after scaling to the results of the maximum-likelihood fit.
The uncertainties reflect the limited number of MC events and the estimated scale factor uncertainties, where appropriate.
The multijet events background contribution is estimated using a data-based procedure. }
\begin{tabular}{R{2.5cm}|d{5}d{5}d{5}}
Process &
\multicolumn{3}{c}{
  \begin{tabular}{C{0.123\linewidth}C{0.1365\linewidth}C{0.17\linewidth}}
$\PQt$ & $\PAQt$ & $\PQt + \PAQt$
  \end{tabular}
}\\
\hline
$\ttbar$ & 1543 , 24 & 1573 , 23 & 3118 , 34 \\
$\PQt\PW$ & 143 , 8 & 168 , 9 & 311 , 12 \\
$s$-channel & 44 , 4 & 27 , 3 & 72 , 4 \\
\hline
$\wjets$ & 1332 , 60 & 1022 , 56 & 2353 , 81 \\
$\zjets$ & 181 , 23 & 189 , 23 & 371 , 32 \\
Diboson & 21 , 2 & 13 , 1 & 33 , 2 \\
\hline
Multijet & 219 , 110 & 208 , 105 & 427 , 214 \\
\hline
$t$-channel & 3852 , 101 & 2202 , 90 & 6049 , 136 \\
\hline
Total expected & 7334 , 165 & 5402 , 153 & 12733 , 271 \\
\hline
Data & 7223 & 5281 & 12504 \\
\end{tabular}
\label{tab:yields}
\end{center}
\end{table}

\section{Unfolding}
\label{sec:unfolding}
An unfolding procedure is used to determine the differential cross section as a function of \costheta\ at the parton level.
It accounts for distortions from detector acceptance, selection efficiencies, imperfect reconstruction of the top quark candidate,
and the approximation made in treating the direction of the untagged jet as the spectator quark direction.

In simulation, the parton-level definition of \costheta\ is defined based on the generated muon from the decay chain of a top quark or antiquark and the spectator quark scattering off the top quark or antiquark via virtual W boson exchange,
with all momenta boosted into the rest frame of the generated top quark or antiquark.
 To preserve the spin information from the W decay, the response matrix takes into account the case in which the muon is from $\mathrm{W}\rightarrow \tau\nu \rightarrow \mu \nu\nu$ decay by unfolding the angular distribution to the $\tau$ lepton.
Prior to unfolding, remaining background contributions are subtracted from the reconstructed data,
using the fitted number of events and their uncertainties, estimated in Section~\ref{sec:fit}.

After the background subtraction, an unfolding procedure~\cite{Blobel:2002pu} is applied. At its core is the application of a matrix inversion using second derivatives for regularisation.
 A detailed description of the procedure can be found in the \ttbar~charge asymmetry analysis~\cite{Chatrchyan:2011hk}, performed previously by CMS, which utilises the same method.

The performance of the unfolding algorithm is checked using sets of pseudo-experiments.
 Pull distributions show no sign that the uncertainties are treated incorrectly.
 A bias test is performed by injecting anomalous $\PW\PQt\PQb$-vertex coupling events as pseudo-data, generated with \COMPHEP~\cite{Boos:2004kh,boos2000}. This test verifies that,
 with the analysis strategy described here, it is possible to measure different asymmetries correctly, and with only a small bias that will be accounted for as a systematic uncertainty.

The value of $A_{\mu}$ is extracted using a $\chi^2$-fit of the unfolded \cosTheta\ distribution,
 under the assumption that Eq.~(\ref{eq:cosThetaDistr}) is valid.  The fit takes into account the bin-by-bin correlations that are induced in the unfolding procedure.

An alternative procedure, based on analytic matrix inversion with only two bins in the \costheta\ distribution
(corresponding to forward- and backward-going muons), is used as a crosscheck.
Although the results of the two methods are in agreement, the expected precision of the analytic matrix inversion is slightly worse when tested using pseudo-data.

\section{Systematic uncertainties}
\label{sec:systematics}
The differential cross section and asymmetry measurement presented in this paper can be affected by several sources of systematic uncertainty. To evaluate the impact of each source, we perform a new background estimation
 and repeat the measurement with systematically shifted
simulated templates and response matrices. The expected systematic uncertainty for each source is
taken to be the maximal shift in the values of the asymmetry between the
nominal asymmetry and the one measured using the shifted templates.

\textbf{ML fit uncertainty}: This uncertainty is determined by propagating the uncertainty associated with the background normalisation from the maximum-likelihood fit through the unfolding procedure.

\textbf{Other background fractions}: A specific uncertainty is assigned to the fraction of each minor process that is combined with similar and larger processes in the fit.
These are dibosons and \zjets\ production for the W/Z/diboson component, and the tW and $s$-channel production for the top quark component.
A yield uncertainty of 50\% is used for each of the templates.

\textbf{Multijet events background shape}: A shape uncertainty is taken into account by varying the range of inverted isolation requirement used
 to extract the templates for estimating this background contribution.

\textbf{Multijet events background yield}: A 50\% uncertainty is assigned to the yield obtained from the multijet events fit.

\textbf{b tagging}: The uncertainties in the b tagging and mistagging efficiencies for individual jets as measured in data~\cite{CMS-PAS-BTV-13-001}
are propagated to the simulation event weights.

\textbf{Detector-related jet and
$\boldsymbol{E_{\mathrm{T}}}\hspace{-1.1em}/\kern0.45em$
effects}: All reconstructed jet four-momenta in simulated events
are changed simultaneously according to the $\eta$- and $\pt$-dependent
uncertainties in the jet energy scale~\cite{Chatrchyan:2011ds}.
 The changes in jet four-momenta are also propagated to $\MET$.
 In addition, the effect on the measurement of $\MET$ arising from the 10\% uncertainty associated with unclustered energy deposits
 in the calorimeters is estimated after subtracting from \MET all jets and leptons. An extra uncertainty accounts for the known difference in JER relative to data~\cite{Chatrchyan:2011ds}.

\textbf{Pileup}: A 5\% uncertainty is applied to the average expected number of pileup interactions in order to estimate the uncertainty arising from the modelling of pileup.

\textbf{Muon trigger, identification, and isolation efficiencies}:
A systematic uncertainty of 1\% is applied independently to the muon trigger, identification, and isolation efficiencies.
 These uncertainties cover the efficiency differences between the phase space regions sampled by the present selection
 and by the selection of \zjets\ events for the tag-and-probe procedure.

\textbf{$\boldsymbol{ \mathrm{ t \bar{t} } }$ top quark
$\boldsymbol{p_{\mathrm{T}}}$
reweighting}: The \MADGRAPH\ model for \ttbar\ production is known to predict a harder \pTtop\ spectrum compared to that observed in data~\cite{Chatrchyan:2012saa,Khachatryan:2015oqa}. Although the correlation with other uncertainty sources is not clear, the spectrum of generator-level top quarks in \ttbar\ events is reweighted to the measured differential cross section and an additional systematic uncertainty from this reweighting by either doubling or not using any reweighting is applied.

\textbf{W boson $\boldsymbol{p_{\mathrm{T}}}$ reweighting in W+jets}: The \MADGRAPH\ model for \wjets\ events predicts a \pT\ spectrum of the reconstructed \PW\ boson candidate that does not agree with data in the ``2 jets 0 tags'' control region.
The distribution is reweighted to data (after subtraction of other processes) and the difference is taken as a systematic uncertainty.

\textbf{$\boldsymbol{\cos{ \theta^{*}_{\mu}} }$ reweighting in W+jets}: The uncertainty associated with the reweighting procedure presented in Section~\ref{sec:wjets}
is estimated conservatively by comparing the result after \cosThetaPol shape reweighting with that determined with no weighting applied.
The difference between the two is then symmetrised and taken as the uncertainty. An additional uncertainty is assigned to the fraction of \wjets events in which jets arise from heavy flavours. This uncertainty is taken into account by scaling its contribution by $\pm50\%$ relative to the prediction by \MADGRAPH.

\textbf{Unfolding bias}: A test of the analysis shows a small bias when injecting events with anomalous couplings as pseudo-data. This is treated as an additional systematic uncertainty in the asymmetry measurement.

\textbf{Generator model}: The nominal result is compared with the one obtained using an unfolding matrix from a signal sample generated with \AMCATNLO,
interfaced with \PYTHIA~8 for parton showering.

\textbf{Top quark mass}: Additional samples of \ttbar and signal events are generated with the top quark mass changed by $\pm 3$~GeV.
 These are used to determine the uncertainty arising from our knowledge of the top quark mass.
 This is a conservative estimate as the current world average is $173.3\pm0.8$~GeV~\cite{ATLAS:2014wva}.

\textbf{Parton distribution functions}:
The uncertainty due to the choice of the set of parton distribution functions (PDF) is estimated by reweighting the simulated events with each of the
52 eigenvectors of the {CT10} collection~\cite{PDF:CTEQ10}, and additional eigenvectors corresponding to variation of the strong coupling,
as well as using the central sets from the {MSTW2008CPdeut}~\cite{MSTW2013} and {NNPDF23}~\cite{Ball:2012cx} collections.
 The {\sc LHAPDF}~\cite{Whalley:2005nh} package is used for the reweighting.

\textbf{Renormalisation and factorisation scales}:
The uncertainties in the renormalisation and factorisation scales
(set to a common scale equal to the momentum transfer $Q$ in the event) are evaluated for signal, \ttbar and \wjets independently,
by doubling or halving the value of the scale.
For the signal, a reweighting procedure is applied to simulated events,
using the simplification of neglecting the scale dependence of the parton shower (PS).
Since the signal process does not contain a QCD vertex at LO in the 5FS, the dependence of its cross section with scale $Q$ can be written as

\begin{equation}
\sigma_{t\mbox{-}\mathrm{ch.}}^\mathrm{LO} (Q)=\int_{0}^{1} \mathrm{d}x_{1} f_\mathrm{PDF}(x_{1},Q^{2})\int_{0}^{1} \mathrm{d}x_{2} f_\mathrm{PDF}(x_{2},Q^{2})\, \hat{\sigma}(x_{1},x_{2}),
\label{eq:ME-scale-dependence}
\end{equation}

where $x_{i}$ are the momentum fractions of the two partons in the colliding protons, $f_\mathrm{PDF}(x_{i},Q^{2})$ is the PDF, and $\hat{\sigma}(x_{1},x_{2})$ denotes the partonic cross section.
 The event reweighting to a different scale $Q^{\prime}$ is then defined using a factor

 \begin{equation}
w_{Q\rightarrow Q^\prime}(x_1,x_2)=\frac{f_\mathrm{PDF}(x_{1},Q^{\prime 2}) \,
f_\mathrm{PDF}(x_{2},Q^{\prime 2})}{f_\mathrm{PDF}(x_{1},Q^{2}) \,  f_\mathrm{PDF}(x_{2},Q^{2})}.
\label{eq:qscale-ME-only-reweighting}
\end{equation}
Dedicated simulated samples with doubled and halved scales are used to verify the validity of the approximation of ignoring the effect of scale in PS simulation for the signal process.
The reweighting is preferred over use of these dedicated samples because of their limited number of events.

For the \ttbar\ and \wjets\ backgrounds, a lower threshold is applied to the \bdtsigbg\ discriminant in simulated samples that have a changed $Q$ scale to increase the number of selected events.
 This provides a \cosTheta\ distribution that agrees, within the limited statistical uncertainty of the simulation,
 with the shape obtained by applying the nominal \bdtsigbg\ discriminant threshold.

\textbf{Matrix element/parton shower matching threshold}:
The impact of the choice of ME/PS matching threshold in the MLM procedure is evaluated independently for \ttbar\ and \wjets processes,
using dedicated samples in which the threshold is either doubled or halved.

\textbf{Limited number of simulated events}: The uncertainty associated with the limited amount of simulated events used in forming the templates
 is taken into account at all stages of the analysis, i.e. both in terms of fluctuations in the background and in determining the elements of the migration matrix.
 The limited number of simulated events can also influence the estimation of other systematic uncertainties, potentially leading to an overestimation of the associated uncertainties.

Table~\ref{tab:syst_combined} shows the impact of the different sources of systematic uncertainties on the asymmetry measurements.

\newcolumntype{.}{D{.}{. }{-1}}

\begin{table}
\renewcommand{\arraystretch}{1.1}
\begin{center}
\topcaption{List of systematic uncertainties and their induced shifts from the nominal measured asymmetry for the top quark ($\delta A_{\Pgm}(\PQt)$), antiquark ($\delta A_{\PGm}(\PAQt)$), and their combination ($\delta A_{\PGm}(\PQt + \PAQt)$).}
\begin{tabular}[htc]{r | .  . . }

\multicolumn{1}{c|}{}
        & \multicolumn{1}{c}{$\phantom{+0}\delta \AmuT / 10^{-2}$}
        & \multicolumn{1}{c}{$\phantom{+0}\delta \AmuTbar / 10^{-2}$}
        & \multicolumn{1}{c}{$\delta \AmuTplusTbar / 10^{-2}$}
         \\

\hline
Statistical & 3.2   & 4.6   & 2.6   \\
\hline
ML fit uncertainty & 0.7   & 1.2   & 0.6   \\
Diboson bkg. fraction & {<}0.1   & {<}0.1   & {<}0.1   \\
\ZJets bkg. fraction & {<}0.1   & {<}0.1   & {<}0.1   \\
$s$-channel bkg. fraction & 0.3   & 0.2   & 0.2   \\
tW bkg. fraction & 0.1   & 0.7   & 0.2   \\
Multijet events shape & 0.5   & 0.7   & 0.5   \\
Multijet events yield & 1.9   & 1.2   & 1.7   \\
\hline
b tagging & 0.7   & 1.2   & 0.9   \\
Mistagging & {<}0.1   & 0.1   & {<}0.1   \\
Jet energy resolution & 2.7   & 1.8   & 2.0   \\
Jet energy scale & 1.3   & 2.6   & 1.1   \\
Unclustered \MET & 1.1   & 3.3   & 1.3   \\
Pileup & 0.3   & 0.2   & 0.2   \\
Lepton identification & {<}0.1   & {<}0.1   & {<}0.1   \\
Lepton isolation & {<}0.1   & {<}0.1   & {<}0.1   \\
Muon trigger efficiency & {<}0.1   & {<}0.1   & {<}0.1   \\
\hline
Top quark \pT reweighting & 0.3   & 0.3   & 0.3   \\
\wjets W boson \pT reweighting & 0.1   & 0.1   & 0.1   \\
\wjets heavy-flavour fraction & 4.7   & 6.2   & 5.3   \\
\wjets light-flavour fraction & {<}0.1   & {<}0.1   & 0.1   \\
\wjets \cosThetaPol reweighting & 2.9   & 3.4   & 3.1   \\
Unfolding bias & 2.5   & 4.2   & 3.1   \\
\hline
Generator model & 1.6   & 3.5   & 0.3   \\
Top quark mass & 1.9   & 2.9   & 1.8   \\
PDF & 0.9   & 1.6   & 1.2   \\
$t$-channel renorm./fact. scales & 0.2   & 0.2   & 0.2   \\
\ttbar renorm./fact. scales & 2.2   & 3.4   & 2.7   \\
\ttbar ME/PS matching & 2.2   & 0.5   & 1.6   \\
\wjets renorm./fact. scales & 3.7   & 4.6   & 4.0   \\
\wjets ME/PS matching & 3.8   & 3.0   & 3.4   \\
\hline
Limited MC events& 2.1   & 3.2   & 1.8   \\
\hline
Total uncertainty & 10.5   & 13.8   & 10.5   \\
\end{tabular}
\label{tab:syst_combined}

\end{center}

\end{table}

\section{Results}
\begin{figure}[th]
\centering
\includegraphics[width=0.49\textwidth]{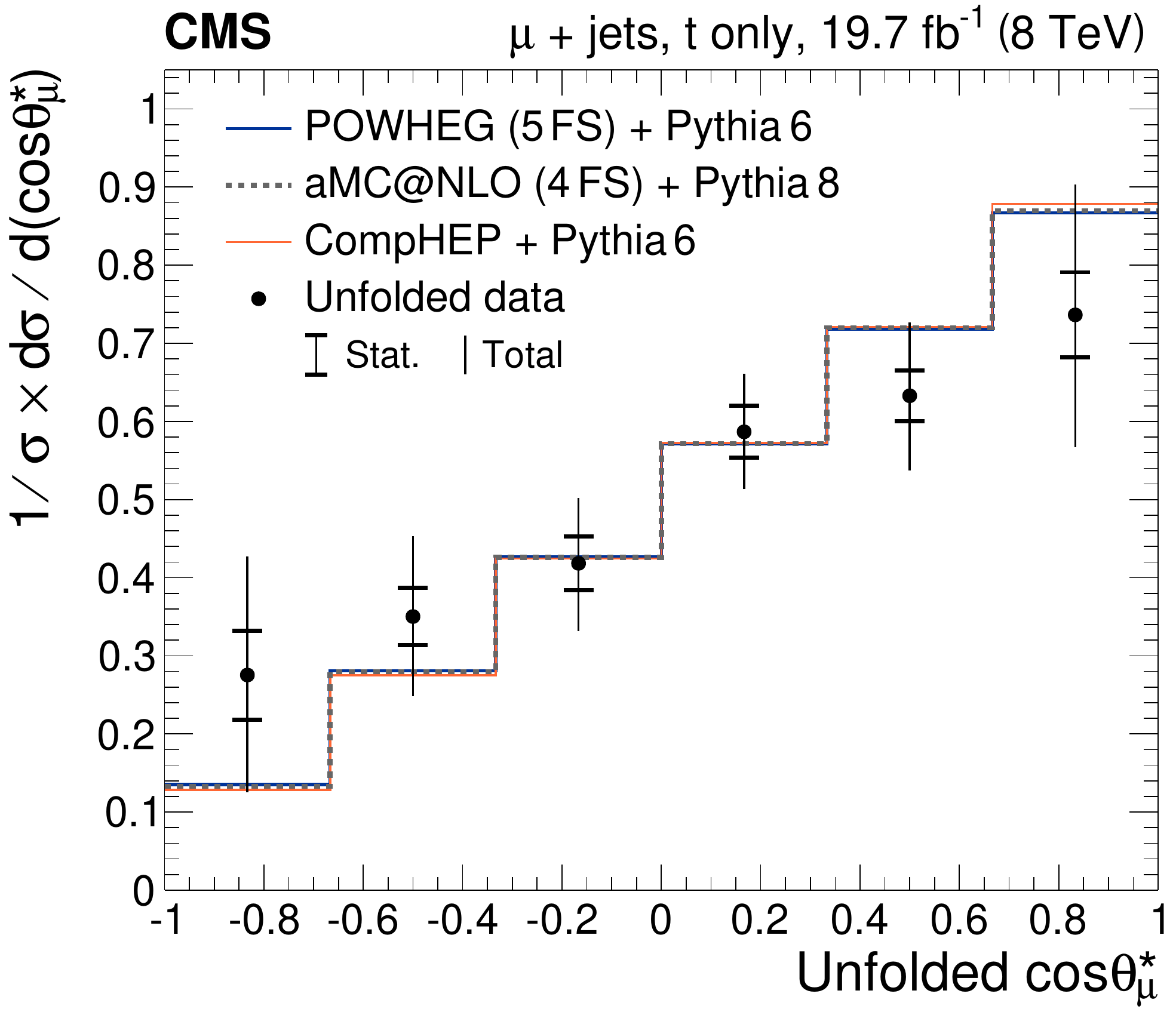}
\includegraphics[width=0.49\textwidth]{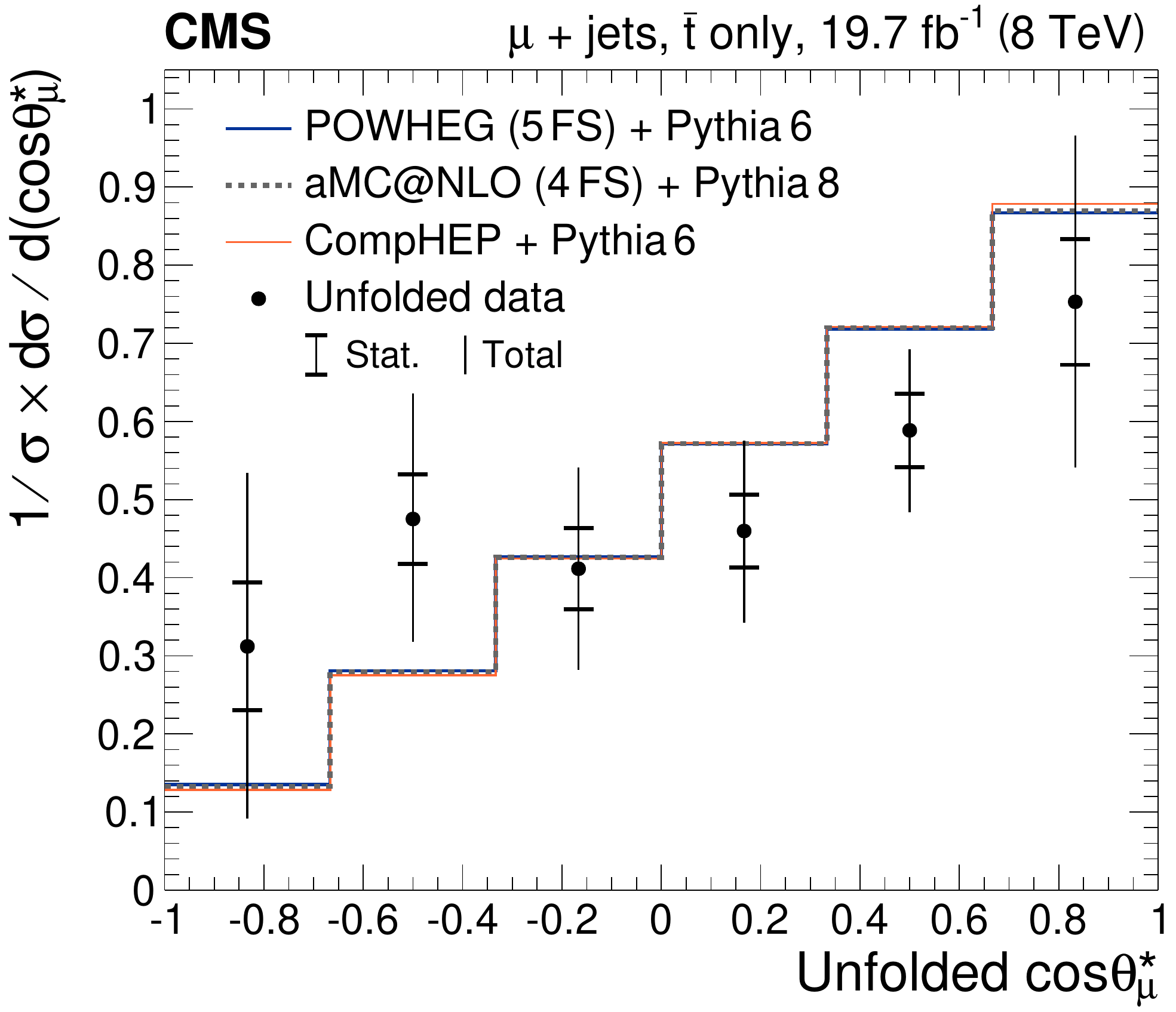}
\caption{The normalised differential cross sections as a function of unfolded \cosThetaPol for (left)~top quark and (right)~antiquark compared to the predictions
from \POWHEG, \AMCATNLO, and \COMPHEP. The inner (outer) bars represent the statistical (total) uncertainties.}
\label{fig:unfolded_charged}
\end{figure}

\begin{figure}[th]
\centering
\includegraphics[width=0.6\textwidth]{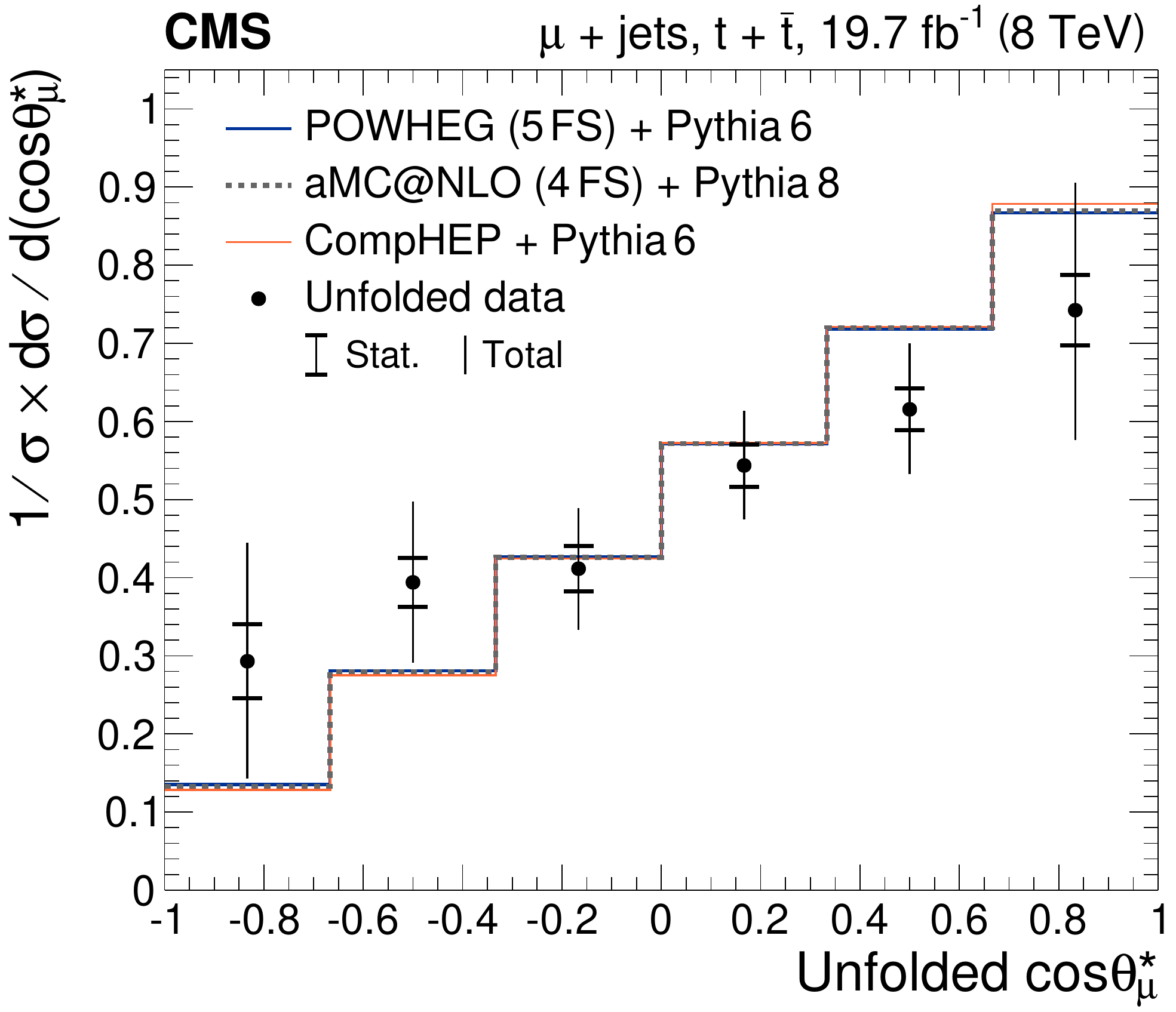}
\caption{The normalised differential cross section as a function of unfolded \cosThetaPol for top quark and antiquark combined,
compared to the predictions from \POWHEG, \AMCATNLO, and \COMPHEP. The inner (outer) bars represent the statistical (total) uncertainties.}
\label{fig:unfolded_comb}
\end{figure}

Figures~\ref{fig:unfolded_charged} and~\ref{fig:unfolded_comb}, respectively,
show the differential cross sections obtained from the unfolding procedure for single top quark and antiquark production, and
for their combination, with a comparison to the SM expectations
from \POWHEG, \AMCATNLO, and \COMPHEP. These generators agree well in their predictions of \Amu.
Uncertainties arising from the renormalisation and factorisation scale and PDF variations have been found to be negligible for the predicted differential distributions and are therefore not shown.

The asymmetry \Amu is extracted from the differential cross section according to Eq.~(\ref{eq:cosThetaDistr}), taking into account correlations.
Using this procedure, we obtain:

\begin{align}
\AmuT          & = \AlResultTopStatSys = \AlResultTop , \\
\AmuTbar       & = \AlResultAntiTopStatSys = \AlResultAntiTop , \\
\AmuTplusTbar & = \AlResultCombinedStatSys = \AlResultCombined ,
\end{align}

where the combined result is compatible with a $p$-value of $\mathrm{\mathit{p}(data|SM)}=\AlResultCombinedPvalue$,
which corresponds to \AlResultCombinedSvalue standard deviations compared to the expected SM asymmetry of $0.44$ as predicted by \POWHEG~(NLO).
Alternatively, the compatibility of the combined result with the hypothetical case of $\Amu =0$ is smaller, yielding a $p$-value of $\mathrm{\mathit{p}(data|}\Amu =0) = 0.7\%$, and corresponding to 2.7 standard deviations.
The SM asymmetry predictions for simulated top quark and antiquark events are equal, while~\cite{Mahlon:1999gz} predicts a $\mathcal{O}(1\%)$ difference, which is small compared to the precision of the current measurement.

As a crosscheck, an analytic 2-bin unfolding is also performed,
which yields the numbers $N(\uparrow)$ and $N(\downarrow)$ defined in Eq.~(\ref{eq:top-quark-asymmetry}).
This gives a compatible but slightly less precise value for \Amu of:

\begin{equation}
\AmuTplusTbar = 0.28\pm 0.03\stat \pm 0.1\syst = 0.28\pm 0.12 .
\end{equation}

\section{Summary}
The first measurement of the top quark spin asymmetry, sensitive to the top quark polarisation,  in $t$-channel single top quark production has been presented. This measurement is based on a sample of pp collisions at a centre-of-mass energy of 8\TeV, corresponding to an integrated luminosity of 19.7\fbinv.

The asymmetry, $A_{\mu}$, is obtained by performing a differential cross section measurement of \cosThetaPol, between forward- and backward-going muons with respect to the direction of the spectator quark in the top quark rest frame.
The measurement yields $A_{\mu} = \AlResultCombinedStatSys = \AlResultCombined$, which is compatible with a $p$-value of $\AlResultCombinedPvalue$, equivalent to \AlResultCombinedSvalue standard deviations, with the standard model expectation.

The asymmetry observed in data is smaller than the prediction. Separate results from exclusive top quark or antiquark events are compatible within the uncertainties.
This difference cannot be explained by any single source of systematic uncertainty considered in this analysis.

\section*{Acknowledgements}

\hyphenation{Bundes-ministerium Forschungs-gemeinschaft Forschungs-zentren} We congratulate our colleagues in the CERN accelerator departments for the excellent performance of the LHC and thank the technical and administrative staffs at CERN and at other CMS institutes for their contributions to the success of the CMS effort. In addition, we gratefully acknowledge the computing centres and personnel of the Worldwide LHC Computing Grid for delivering so effectively the computing infrastructure essential to our analyses. Finally, we acknowledge the enduring support for the construction and operation of the LHC and the CMS detector provided by the following funding agencies: the Austrian Federal Ministry of Science, Research and Economy and the Austrian Science Fund; the Belgian Fonds de la Recherche Scientifique, and Fonds voor Wetenschappelijk Onderzoek; the Brazilian Funding Agencies (CNPq, CAPES, FAPERJ, and FAPESP); the Bulgarian Ministry of Education and Science; CERN; the Chinese Academy of Sciences, Ministry of Science and Technology, and National Natural Science Foundation of China; the Colombian Funding Agency (COLCIENCIAS); the Croatian Ministry of Science, Education and Sport, and the Croatian Science Foundation; the Research Promotion Foundation, Cyprus; the Ministry of Education and Research, Estonian Research Council via IUT23-4 and IUT23-6 and European Regional Development Fund, Estonia; the Academy of Finland, Finnish Ministry of Education and Culture, and Helsinki Institute of Physics; the Institut National de Physique Nucl\'eaire et de Physique des Particules~/~CNRS, and Commissariat \`a l'\'Energie Atomique et aux \'Energies Alternatives~/~CEA, France; the Bundesministerium f\"ur Bildung und Forschung, Deutsche Forschungsgemeinschaft, and Helmholtz-Gemeinschaft Deutscher Forschungszentren, Germany; the General Secretariat for Research and Technology, Greece; the National Scientific Research Foundation, and National Innovation Office, Hungary; the Department of Atomic Energy and the Department of Science and Technology, India; the Institute for Studies in Theoretical Physics and Mathematics, Iran; the Science Foundation, Ireland; the Istituto Nazionale di Fisica Nucleare, Italy; the Ministry of Science, ICT and Future Planning, and National Research Foundation (NRF), Republic of Korea; the Lithuanian Academy of Sciences; the Ministry of Education, and University of Malaya (Malaysia); the Mexican Funding Agencies (CINVESTAV, CONACYT, SEP, and UASLP-FAI); the Ministry of Business, Innovation and Employment, New Zealand; the Pakistan Atomic Energy Commission; the Ministry of Science and Higher Education and the National Science Centre, Poland; the Funda\c{c}\~ao para a Ci\^encia e a Tecnologia, Portugal; JINR, Dubna; the Ministry of Education and Science of the Russian Federation, the Federal Agency of Atomic Energy of the Russian Federation, Russian Academy of Sciences, and the Russian Foundation for Basic Research; the Ministry of Education, Science and Technological Development of Serbia; the Secretar\'{\i}a de Estado de Investigaci\'on, Desarrollo e Innovaci\'on and Programa Consolider-Ingenio 2010, Spain; the Swiss Funding Agencies (ETH Board, ETH Zurich, PSI, SNF, UniZH, Canton Zurich, and SER); the Ministry of Science and Technology, Taipei; the Thailand Center of Excellence in Physics, the Institute for the Promotion of Teaching Science and Technology of Thailand, Special Task Force for Activating Research and the National Science and Technology Development Agency of Thailand; the Scientific and Technical Research Council of Turkey, and Turkish Atomic Energy Authority; the National Academy of Sciences of Ukraine, and State Fund for Fundamental Researches, Ukraine; the Science and Technology Facilities Council, UK; the US Department of Energy, and the US National Science Foundation.

Individuals have received support from the Marie-Curie programme and the European Research Council and EPLANET (European Union); the Leventis Foundation; the A. P. Sloan Foundation; the Alexander von Humboldt Foundation; the Belgian Federal Science Policy Office; the Fonds pour la Formation \`a la Recherche dans l'Industrie et dans l'Agriculture (FRIA-Belgium); the Agentschap voor Innovatie door Wetenschap en Technologie (IWT-Belgium); the Ministry of Education, Youth and Sports (MEYS) of the Czech Republic; the Council of Science and Industrial Research, India; the HOMING PLUS programme of the Foundation for Polish Science, cofinanced from European Union, Regional Development Fund; the OPUS programme of the National Science Center (Poland); the Compagnia di San Paolo (Torino); MIUR project 20108T4XTM (Italy); the Thalis and Aristeia programmes cofinanced by EU-ESF and the Greek NSRF; the National Priorities Research Program by Qatar National Research Fund; the Rachadapisek Sompot Fund for Postdoctoral Fellowship, Chulalongkorn University (Thailand); and the Welch Foundation, contract C-1845.

\bibliography{auto_generated}

\cleardoublepage \appendix\section{The CMS Collaboration \label{app:collab}}\begin{sloppypar}\hyphenpenalty=5000\widowpenalty=500\clubpenalty=5000\input{TOP-13-001-authorlist.tex}\end{sloppypar}
\end{document}

%% file: TOP-13-001-authorlist.tex
\textbf{Yerevan Physics Institute,  Yerevan,  Armenia}\\*[0pt]
V.~Khachatryan, A.M.~Sirunyan, A.~Tumasyan
\vskip\cmsinstskip
\textbf{Institut f\"{u}r Hochenergiephysik der OeAW,  Wien,  Austria}\\*[0pt]
W.~Adam, E.~Asilar, T.~Bergauer, J.~Brandstetter, E.~Brondolin, M.~Dragicevic, J.~Er\"{o}, M.~Flechl, M.~Friedl, R.~Fr\"{u}hwirth\cmsAuthorMark{1}, V.M.~Ghete, C.~Hartl, N.~H\"{o}rmann, J.~Hrubec, M.~Jeitler\cmsAuthorMark{1}, V.~Kn\"{u}nz, A.~K\"{o}nig, M.~Krammer\cmsAuthorMark{1}, I.~Kr\"{a}tschmer, D.~Liko, T.~Matsushita, I.~Mikulec, D.~Rabady\cmsAuthorMark{2}, B.~Rahbaran, H.~Rohringer, J.~Schieck\cmsAuthorMark{1}, R.~Sch\"{o}fbeck, J.~Strauss, W.~Treberer-Treberspurg, W.~Waltenberger, C.-E.~Wulz\cmsAuthorMark{1}
% ...
\vskip\cmsinstskip
% ...
1:~~Also at Vienna University of Technology, Vienna, Austria\\
2:~~Also at CERN, European Organization for Nuclear Research, Geneva, Switzerland\\
% ...
